%% file: paper.tex
\documentclass{llncs}

\usepackage{graphicx}
\usepackage{hyperref}
\usepackage{xcolor}
\usepackage{amsmath}
\usepackage{ltablex}
\usepackage{algorithm}
\usepackage{algpseudocode}
\usepackage{wrapfig}
\usepackage[title]{appendix}

\usepackage{subcaption}
\usepackage{multirow}
\usepackage{array}

\makeatletter
\newcommand{\algmargin}{\the\ALG@thistlm}
\makeatother
\newlength{\whilewidth}
\algdef{SE}[parWHILE]{parWhile}{EndparWhile}[1]
  {\parbox[t]{\dimexpr\linewidth-\algmargin}{%
     \hangindent\whilewidth\strut\algorithmicwhile\ #1\ \algorithmicdo\strut}}{\algorithmicend\ \algorithmicwhile}%
\algnewcommand{\parState}[1]{\State%
  \parbox[t]{\dimexpr\columnwidth-\algmargin}{\strut #1\strut}}

\algnewcommand{\LineComment}[1]{\State \(\triangleright\) #1}
\newcommand{\paragraphb}[1]{\vspace{0.03in}\noindent{\bf #1} }

\begin{document}
\title{Secure Calibration for Safety-Critical IoT: Traceability for Safety Resilience}
%
\author{Ryan Shah$^\dagger$, Michael McIntee$^\dagger$, Shishir
  Nagaraja$^\dagger$, Sahil Bhandary$^\ddag$, Prerna Arote$^\ddag$,
  Joy Kuri$^\ddag$}
\institute{The University of Strathclyde$^\dagger$, Indian Institute of Science$^\ddag$}

\maketitle

\begin{abstract}
  Secure sensor calibration constitutes a foundational step that
  underpins operational safety in the Industrial Internet of
  Things.
  While much attention has been given to IoT security such as
  the use of TLS to secure sensed data, little thought has been given to
  securing the calibration infrastructure itself. Currently traceability
  is achieved via manual verification using paper-based datasheets which
  is both time consuming and insecure. For instance, when the
  calibration status of parent devices is revoked as mistakes or
  mischance is detected, calibrated devices are not updated until
  the next calibration cycle, leaving much of the calibration parameters
  invalid. Aside from error, any party within the calibration
  infrastructure can maliciously introduce errors since the current
  paper based system lacks authentication as well as non-repudiation. In
  this paper, we propose a novel resilient architecture for calibration
  infrastructure, where the calibration status of sensor elements can be
  verified on-the-fly to the root of trust preserving the properties of
  authentication and non-repudiation. We propose an implementation based
  on smart contracts on the Ethereum network. Our evaluation shows that
  Ethereum is likely to address the protection requirements of traceable
  measurements.
\end{abstract}

\input{sections/introduction}

\input{sections/background}
\input{sections/protection_requirements}
\input{sections/evaluation}
\input{sections/discussion_ryan}
\input{sections/conclusion}

\section{Acknowledgements}
\label{sec:ack}

The authors are grateful for the support by Engineering and Physical Sciences Research Council (11288S170484-102), National Physical Laboratory, Keysight Inc (6017), UKIERI-2018-19-005, and the Department of Science and Technology (DST), Govt. of India.

\bibliographystyle{splncs04}
\bibliography{references}

\end{document}

%% file: sections/introduction.tex
\section{Introduction}
\label{sec:introduction}

IoT cyberphysical systems, such as connected robots, are increasingly
transforming a wide range of application areas, including but not limited
to surgical suites~\cite{talamini2003prospective} and industrial processing
plants~\cite{quarta2017experimental}. The use of automation in these areas
brings forth the potential to increase the efficiency of output, yet
accuracy and precision under adversarial pressure remains a constant
worry. In the context of surgical robotics, for example, a high degree
of accuracy and precision must be maintained as accurate sensing could
mean the difference between life and death.

A safety-critical device such as a surgical
robot~\cite{pereira2016uncertainty} involves much more work. A
calibration infrastructure distributed across the OEMs,
third-party calibration agencies and suppliers involved in the supply
chain~\cite{huang2005computer,alicke2017supply}, are part of the
calibration work-flow. The root of trust (calibration integrity) is a
National Measurement Institute (NMI) that maintains the gold standards
for sensing and measurment. This is typically a government agency such
as the National Physical Laboratory (NPL) in UK or the NIST in
USA. The root of trust for each type of sensor, consists of a master
calibration device, which is used to calibrated other reference devices
that serve as a proxy for the master, and are in turn used to keep
calibration units closer to the field calibrated.

First, as we start to rely on connected robots to perform critical tasks, we
will start to see at least three changes. The security of the
calibration infrastructure itself will start gaining importance and
mechanisms will be required to deal with attackers in the calibration
ecosystem.

Second, calibration correctness becomes a safety-critical
requirement. We argue that the way
ahead, is to ensure that all sensed data is subject to verification
via {\em on-the-fly calibration checks}. This notion involves tracing
sensed measurements to the corresponding gold standard, by involving
all stakeholders: from the operator (e.g. surgeon in a hospital), to
the manufacturer and their suppliers.

Third, how can the operator, regulator, manufacturer, and calibration
agencies work together to create a tamper-resistant trail of recorded
activity to aid system forensics, which can withstand hostile scrutiny
in a court of law when things go wrong? There have been cases of
lawsuits filed by patients, accusing hospitals of negligence over
safety considerations when surgical robots have inflicted accidental
injuries~\cite{lawsuitsurgical,davincilawsuit}, and such are illustrative of the significant liabilities and
stakes involved when ensuring robot safety.

Ultimately, calibration activity -- such as end-to-end measurement and calibration
traceability~\cite{jcgm2012200,kaarls1997comite,de2000calibration,lukavc2015fourth,roblek2016complex,lu2017industry}
-- move from a one-off to a continuous or periodic process, to minimise
errors arising from calibration and associated liabilities. Ensuring the correctness
of calibration in the face of attackers is a crucial requirement that must
be enforced to address the operational resilience requirements of connected
systems.

%% file: sections/background.tex
\section{Background}
\label{sec:background}

To ensure measuring instruments provide high quality and accurate measurements, we must ensure that they are calibrated against a trustworthy source. All measurements have a quantifiable degree of uncertainty and the challenge is to ensure that we can minimise this uncertainty, while maintaining a quantifiable indication of the quality of measurement. National standards for weights and measures are maintained by National Measurement Institutes (NMIs), such as the National Physical Laboratory (NPL) in the United Kingdom. NMIs define national measurement standards, which are associated with values of uncertainty and are used to calibrate measuring instruments.

\begin{figure}[h]
  \centering
  \includegraphics[width=.8\columnwidth]{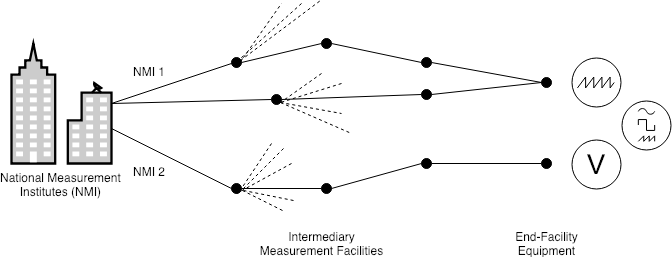}
  \caption{Traceability Chain Paths}
  \label{fig:tpath}
\end{figure}

The calibration of measuring instruments ensures that recorded measurements are of high quality and accuracy, such that they are compared to a standard of higher accuracy to identify errors in instrument readings. We calibrate to meet quality audit requirements and ensure reference designs, subsystems and integrated systems perform as intended. A reliable measurement should be recorded by instruments with low measurement uncertainty and is traceable to corresponding SI units, to a standard or reference method~\cite{de2000calibration}. Traceability is at the heart of measurements and is a basis for comparisons against valid measurements. A measurement's metrological traceability is its property, such that the measurement result is related to a stated reference, through an unbroken chain of calibrations~\cite{jcgm2012200}. Shown in Figure~\ref{fig:tpath} are paths in a traceability chain. As demonstrated by the diagram, each piece of end user equipment, hereafter referred to as an end node, can be traced back along the path to intermediary measurement facilities and ultimately to NMIs --- which refer to the SI units as the basis for calibration. Each node in a path, being an NMI or intermediary facility, can branch out to other intermediary or end nodes, such that each piece of equipment can be used to calibrate a number of others.

Kaarls and Quinn state that a set of defined standard, or reference, methods can be created such that primary method(s) are used to validate or calibrate secondary or tertiary methods, which can be linked to a working-level method~\cite{kaarls1997comite}. The use of primary methods are often time consuming and costly. A trade-off for typical working-level methods induce simplicity, but increases uncertainty. de Castro et al. state that measurement uncertainty is an operationally defined method of detailing the level of confidence associated with a measurement~\cite{de2000calibration}, offering advantages over other terms such as precision and trueness.

\if 1

When an instrument is calibrated, a  calibration report is typically produced, along with a measurement report, which outlines instrument errors, corrections to be applied and the uncertainties in these. Some organisations who require their instruments to be calibrated might use in-house or external measurement facilities to calibrate their instruments. These facilities will calibrate instruments to their standards and their calibration equipment will also be calibrated to some standard. This chain of calibrations ends at a national standards body, such as NPL. In it's simplest form, measurements are still recorded and can be traced over time. Through calibration and measurement uncertainty, we are able to quantitatively assess the quality of a result. Throughout the traceability chain, uncertainties and calibration corrections must be applied and propagated properly to ensure the correct treatment of uncertainties.

\fi

%% file: sections/protection_requirements.tex
\section{Protection requirements}
\label{sec:requirements}

Having noted that record-keeping around the calibration process is a foundational challenge to high-assurance IoT systems, several questions arise. What new security problems and what protection opportunities arise where the typical factory may have upwards of a 100,000 sensors, and thousands of such factories or labs share a few hundred calibration facilities. Clearly a future calibration framework will have to ensure good separation between rivals while also supporting dependable shared channels to ensure traceability chains back to a root calibration unit. If some of the calibration units are left on client facilities then are themselves susceptible to occasional compromise.

The primary asset here are the calibration records i.e the calibration information linking parents to children, along with supporting information to determine various measurement uncertainties such as the offsets to be applied as  environmental parameters vary. Tampering with calibration records would likely be the simplest attack vector to attacking IoT devices at scale.

A capable insider might be able to compromise a subset of record storage perhaps as a consequence of a targeted attack on storage servers. We have to, therefore, design the storage infrastructure such that even if it comes under a massive service denial attack or other form of partial compromise, that traceability, uncertainty, or calibration-related checks should not be impacted. This is rather important. The denial or delay of calibration verification ultimately has a direct bearing on the operational efficiency -- a surgery involving a surgical-robot cannot commence unless the robot's calibration status has been thorougly verified at the point of use, in order to discharge the operator of liabilities arising from the use of miscalibrated devices.

This brings us to the next requirement. The operator must be able to
{\em prove} that the calibration status of the robot is fully
traceable to national standards to the hospital's insurer in a
publicly verifiable manner. A proof means there is no uncertainty
about the calibration status of the robot prior to surgery. Current
paper-based records cannot be revoked remotely. Digital calibration
reports on the other hand are routinely revoked whenever any
intelligence about the reliability of parent devices anywhere on the
chain emerges. Validating the freshness of calibration reports, and
provably so is therefore of crucial importance in safety-critical IoT
systems. If the proof is publicly verifiable, i.e without the
possession of special credentials or keys, then the system also has
the {\em transparency property}. Transparency refers to the capability
that users can challenge and verify the claimed calibration status of
a device. Transparency enables downstream users to check the integrity
of the data supply chain. We argue that transparency is an essential
requirement for safety-critical systems.

Scale is another important factor that designers must consider. As we
move from the current deployment of networks involving a few connected
sensors to safety-critical systems with hundreds of thousands of
sensors, we will need a calibration verification infrastructure that
scales to the internet.

Anonymity is also an important requirement to protect the privacy of
the data supply chain. Users who verify device calibration risk
exposing themselves, as device users, to the calibration-verification
infrastructure (run by manufacturers). Information can leak as a
result of calibration and traceability checks, as checks expose user
workflows, which can possibly break confidentiality by linking the
verifier to the device. This compromises the privacy of the data
supply chain, as the user's workflow increasingly become visible to
the calibration infrastructure. To ensure that the calibration
infrastructure does not learn any information about the workflow of
specific users, verifier-side (user side) anonymity is an important
requirement.

A summary of the protection requirements described are as follows: enabling collaboration between operators, manufacturers, regulators, calibration agencies via  storing calibration records in a tamper-evident manner in a fully decentralised manner (absence of a trusted third-party); allow operators to transparently verify and prove calibration-status prior to device usage; ensuring operator's metadata (calibration-verification traffic) is protected from traffic analysis attacks via suitable anonymous interfaces.

\subsection{Threat model}
The current verification process for calibration information has no associated
threat model and thus to enable the need for digitisation, a sound threat
model is the first step towards resilience. We believe
there are at least four types of threats to the calibration
infrastructure.

{\em Large-scale compromise:} First, an intentional attack by a state
or state-sponsored group could discover systemic weaknesses that
compromises a large fraction of the calibration infrastructure. These
vulnerabilities could be exploited by a capable attacker resulting in
seeding significant confusion in the best case. And, in the worst-case
scenario, entire batches of a production-cycle might be compromised
such as a whole batch of wrongly proportioned paracetamol landing up
on a supermarket shelf.

{\em Behavioural economics:} Second, as the digital calibration
infrastructure develops into hierarchical trees of substantial size
with millions of participants, complex behaviours may arise as a
result of system economics. For instance, selfish behaviours may
manifest that optimises the costs of a fraction of the participant at
the expense of the rest of the calibration ecosystem.

{\em Flying debris:} Third, secondary impacts of attacks directed at
other targets may damage the calibration infrastructure. For example,
a DoS attack may cause verification to fail if the network is shared
with other systems. If verification is substantially delayed,
it could make instruments uncontrollable triggering a precautionary
shutdown.

{\em Insider threat:} Fourth, an insider may sabotage the calibration
verification infrastructure. For instance, by inducing errors into
agents responsible for storing calibration reports or those
responsible for ensuring functional-semantics over verification
processes. We exclude insiders from physical attacks on calibrated
instruments/devices from the scope of this paper, as as the focus of
this work, is on storage, processing, and verification.

\subsection{Security policy}
Following the threat model, the next step is to develop a security
policy for the calibration system. A security policy is a succinct
description of information flow constraints that stipulates the
protection requirements to be met by security mechanisms, in order to
mitigate the threats outlined in the threat model. Information flow
controls are important. A move from the current peer-to-peer
architecture underlying calibration devices and field instruments, any
of which will cause havoc if compromised, can bring real benefits. The
natural hierarchy within the calibration infrastructure when composed
with information flow controls can compartmentalise risk, thus the
compromise of a few units will do no more than local damage.

We argue that the appropriate information flow control for a
calibration system is multi-level integrity, with root-calibration units
calibrated by primary methods and references at the upper levels, field
devices calibrated by secondary methods and references at the middle levels,
and working level methods, references, and end-user equipment situated
at the bottom. Also known as the BIBA model, this is similar to
multi-level security systems typically used by government systems to
enforce confidentiality by allowing information to flow from
low-confidentiality to high-confidentiality levels
(Eg. Top-secret to Secret to Confidential to Unclassified).


In the proposed calibration security-policy model, information flows
downwards from the root-calibration units that implement the highest
accuracy of measurement, to the measurement level composed of field
devices, and finally to the monitoring system. There is also some
compartmentation at the measurement level (e.g. separate virtual
networks for unrelated near-root calibration units) and even more
compartmentation at the top at the calibration level (where a
gold-standard or other root-level resource is protected). Within the
calibration infrastructure there will be typically three or so
levels. First, the (root or) near-root calibration instruments will
typically perform calibration operations on a field device. After
operation, the calibration-devices in question notify the monitoring
system which lies at the third level. Measurement itself takes place
at the second level where field devices operate. The functionality of
any resource at the calibration-level should never be compromised by
anything that happens at the measurement level; in other words,
information flows from the calibration (unit) to measurement
(instrument) but not vice versa. The monitoring exists at the third
level and is composed of external monitoring systems that take a data
feed from the calibration and measurement levels but that report
externally, for example to give the manufacturer data for consolidated
event reporting, measurement quality, stability monitoring, line fault
location, asset condition and calibration code compliance. Such
systems should not pass data directly up to the calibration system.
Thus we have a multilevel system in which we want to prevent
information flows from monitoring up to measurement and from
measurement up to calibration.

Information-flow constraints across the multiple levels of integrity
can be enforced by a firewall appliance located close to the routing
infrastructure such as a hardware-based firewall appliance
(like Juniper SSG/SRX) which runs on hardware tailored to install as a
network device. This provides enough network interfaces and CPU to
protect small networks (few ports and few Mbps throughput) to
enterprise-level networks (tens of ports, Gbps throughput).
Aside from the previously mentioned
rules, the firewall should further control information flows in order
to prevent the IP addresses of measurements instruments becoming
externally reachable or visible to the outside world. The firewall
should default deny. The default position when configuring the firewall
should be to deny traffic, not permit it. Then information flows can be
controlled accordingly on a whitelist basis. To prevent
side effects of attacks targeting external systems causing
service-denial or worse to the calibration infrastructure. Likewise,
the firewall should prevent the calibration infrastructure from being
used as a stepping stone for attacks on external systems.

The proposed security architecture for sensor and device calibration
is as illustrated in Figure~\ref{fig:calibrationhierarchy}, has upper
levels which consists of root-calibration units operated and managed
by NMIs such as the National Physical Laboratory, each of which
calibrate and manage the accuracy of tens of level 1 calibration
devices, and each level 1 device in turn manages a few thousand level
2 calibration devices, each of which manage the calibration of tens of
thousands of field level instruments. By coupling the calibration
hierarchy with information-flow constraints, we can organise the
measurement infrastructure so that only the compromise of top-level
calibration devices can cause erroneous measurement at scale, thus
reducing the number of critical components at least by a factor of
hundred.  Furthermore, with the use of appropriate controls at level
2, the compromise of a level 2 calibration device does little damage
outside of its first-hop neighbours, then we can arrange to further
reduce the sites of critical failure by another factor of ten. The
calibration hierarchy can be readily extended, without much
imagination, to map the hierarchical levels to local calibration
components within a manufacturing environment.

We assume the root (level 0) and level 1 calibration devices can
(rarely) suffer accidental configuration errors but are otherwise
trustworthy. On the other hand, level 2 calibration devices may suffer
occasional compromise and, as previously mentioned, a fraction of
field instruments may be compromised at any one time which might
misbehave, intentionally or otherwise.

\subsection{Calibration levels and Measurement levels}


We expect that most of the calibration work will be carried out by one or more
middle levels consisting of mid-level calibration devices and calibrated
field instruments that exist in the middle level. Level 1 and 2 organisations
generally use master calibration units to calibrate other devices.
This allows other calibration devices to be calibrated locally,
reducing the time and cost for calibrating at level 0, as only the master unit
needs to be sent to the level 0 organisation. Field devices are sent to
level 1 and 2 organisations for calibration.

On the other hand, all the measurement work will be carried out by the
field devices located at the bottom (leaf position) of the hierarchy.

{\em Let them out, but not in:} As previously described, any close-to-field
calibration devices and field instruments may become a point of compromise
at any one time, causing them to misbehave, intentionally or otherwise.
A newly established network architecture, to define and constrain the
behaviours --- whether malicious or legitimate --- of field instruments and
close-to-field calibration devices, could invoke the use of refusing incoming
connections and only allowing outgoing connections. Field instruments and
close-to-field devices will be primarily used to transmit outgoing data
and not receive incoming data.

{\em Enforcing non-repudiation:} As well as constraining the behaviour of
field instruments and close-to-field devices, a discussion of mitigating
possible compromise is necessary. An important point for mitigation is
to ensure that instruments and devices are accountable for transmitted data,
such as measurements field instruments may take and results from calibration
units. The data should be recorded such that it can be traced back to the
unit itself. This aids in the isolation of a device in the event of compromise.

{\em Access granted:} Across factory premises and different sites, what
shared and private states are practical to hold and will any limitations
be imposed as a result of state? A suitable access control policy should
be defined such that calibration information can be made public by default,
with organisations enabling an option to not publicly display this if they
consider the information to be private. However, the discussion of privatising
calibration information imposes a degree of difficulty on enabling the
traceability of measurements associated with the privatised information.
Therefore, to aid in reducing the difficulty of this process, the calibration
framework could support an anonymised  base system which also allows
revocation. A set of scopes can be defined for the nodes in the traceability
chain as shown in Figure~\ref{fig:tpath}, which determine the access
constraints for data contained within the scope, whilst a general access policy
can be used to cover data in a general scope.

\subsection{Monitoring}
Monitoring is a logical service in the network. The purpose of
monitoring is to collect statistics from both calibration and
measurement levels. Monitoring makes available its information to
relevant users and operators so they can watch and intervene if
needed. This service can perform both passive and active
monitoring. Passively, it can measure statistics such as the number of
measurements that match a certain pattern, the extent of traceability
up the calibration hierarchy, or per-instrument error
margins. Actively, it can interrogate a field instrument by sending a
measurement request and observe the the instrument output. Monitoring
also exposes a new level of control to the calibration
infrastructure. The potential of using this for auditing and
information flow analysis is immense. Among others, this makes
available an interesting potential for tackling malware outbreaks as
well as adapting and reacting to other forms of network attacks. The
monitoring level also feeds data back into the measurement level.

The monitoring service knows all the calibration traffic that
arises from traceability checks issued from the top such as discarding
the use of measurements that aren't traceable to a calibration device
in the top-level, it can check if they are followed, completing the
loop. It can can also actively generate 'test' measurement traffic to
isolate devices that are dishonest. Most importantly, it links the
operators of field devices with the rest of the calibration
infrastructure. While the level 0 devices are the technical
authorities in calibration, the measurement service translates human
operator intentions into control directives.  The monitoring service
exposes a interface for a users of the measurement network, or to
those to whom the operator delegates access, for example an operator
for a site should not be able to access measurement data from other
departments or manufacturing sites even if they share a common
calibration infrastructure, for instance to restrict the frequency and
outputs of measurement probes operated by Shell and BP. Each such user
gets a separate slice of the measurement service along with relevant
devices, traceability information, and monitoring data. On the
management side, the interface deals with resource allocation and
visibility; and the monitoring side only shows the part of the
collected data that the user is authorised to view.

\begin{figure}[ht!]
 \centering
  	\begin{subfigure}{.5\textwidth}
      \centering
      \includegraphics[width=0.5\columnwidth]{img/calcert}
      \caption{Example Calibration Report}
      \label{fig:calcert}
	\end{subfigure}%
	\begin{subfigure}{.5\textwidth}
    \centering
    \includegraphics[width=0.6\columnwidth]{img/signing}
    \caption{Signing Calibration Certificates}
    \label{fig:signing}
	\end{subfigure}
\end{figure}

\subsection{Protection mechanism}
\label{sec:protmech}
Given the protection requirements, it is natural to pursue the development of a safety assurance system which enables data traceability and support for system forensics -- in comparison with current state-of-the-art calibration systems~\cite{leach1999calibration,hackel2017digital}.
In order to provide decentralised tamper-evident storage, protect
operator anonymity, and enable transparency at the scale of the
Internet requires a highly distributed infrastructure that does not
have any centralised components. These requirements fit the blockchain
as a solution very well. In accordance with our protection
requirements for maintaining high integrity and tamper-resistance, a
blockchain uses strong cryptographic links among blocks, as well as a
distributed network for storage and consensus, making it extremely
hard to tamper with or delete data from the blockchain-based
calibration datastore. This not only aids in fulfilling our integrity
requirement, but also enforces non-repudiation. Since the blockchain
is a ledger, keeping records of all transactions, we can ensure that
calibration devices at all levels of the hierarchy cannot deny
calibration operations or calibration report, and can thus be held
accountable for their actions.

Instead of a centralized trusted authority to store calibration
records, a peer to peer network of nodes independently maintains the
entire database, with cryptographic algorithms used to secure data
integrity. This is a blockchain~\cite{wood2014ethereum}. Each node receives
transactions in a different order. All transactions received within a
certain time frame are aggregated into a new block. To
ensure that the data in every block is consistent across all nodes, a
voting scheme is used, which is called mining. Instead of having a
direct vote, a proof-of-work model is used to determine which node
wins the vote to determine the next accepted block. Each node first
aggregates a limited number of transactions, and then solves an NP
cryptographic problem. This problem has been formulated such that any
random node may stumble upon the solution, but the entire network will
solve the problem given a certain amount of time, which is determined
by the block difficulty.

To provide an incentive for nodes to behave honestly, an incentive
system is needed so that each node processes the transactions it
receives into the next block. Hence each transaction also has an
associated transaction fee which is paid to the node that mines the
block which includes the transaction. Thus, each node is incentivized
to solve as many blocks as possible. This also has the additional
effect on the nodes causing them to balance their workload between
mining a block and processing transactions. The time to mine a block
is set by adjusting the block difficulty, while the transaction
processing time is limited by the transaction limit. In case of simple
transactions like in bitcoin, the transaction limit is enforced by the
total number of transactions. In Ethereum (the mechanism we use), the
transaction varies based on the cost (gas) of the
transaction~\cite{wood2014ethereum}.  Each transaction
consists of performing certain fixed operations defined by the smart
contract, and finally modifying the persistent data on the
blockchain. Instead of keeping the transaction limit as a simple
multiple of transactions, the total cost is instead the sum of cost of
each individual transaction. This limit ensures that the total
computational resources used by the node to process all transactions
remains below a fixed value. Thus, even if the sum of cost of all
transactions exceeds the limit, the limit is determined in such a way
that the time each node takes to process a transaction is far lesser
than the time allocated to solve for proof-of-work. This ensures that
any transaction whose cost is less than the block limit will be
executed in a single block itself, making its round time equal to the
block processing time, regardless of the cost of the actual
transaction.

First, we must consider what will be stored on the blockchain to aid
with functions such as verifying the completeness of traceability
chains in order to accept valid measurements, as well as providing a
way to trace measurements back to field devices. From the calibration
hierarchy, we know that all devices and units are associated with a
calibration report, which outlines information about parent
calibration units, operating ranges with a measurement uncertainty
(MU), among other things. Figure~\ref{fig:calcert} depicts an example
calibration report. As well as this, the report will also detail the
calibration technician who performed the calibration on the device or
unit. Therefore, for completeness, storing reports as well as
technicians on the chain is ideal. This will enable the contract to
verify the calibration status of each device by looking up its
associated parent unit(s), to trace upwards to the master calibration
device (root) unit, which enables
the device user to check whether the device is calibrated against the
root units that establish the gold standard. The use of ECDSA
signatures prevents an adversary from forging calibration reports into
the blockchain (explained in detail below). Also, to prevent the
unauthorised use of valid calibration devices, the traceability-check
contract verifies the signature of technicians all along the
calibration hierarchy. A valid technician's signing keys must be
signed by the calibration organisation's root signing key, and in turn
signed by the NMI, which is the root of trust.

Second, considering the trace back to the calibration technician, we would also want to know what organisation certified the technician to perform calibration, and therefore we must also store the organisations in the calibration hierarchy, to allow for complete audit trails in the event of disaster which stems from invalid or improper calibration.

\begin{figure}[h!]
  \centering
  \includegraphics[width=0.6\columnwidth]{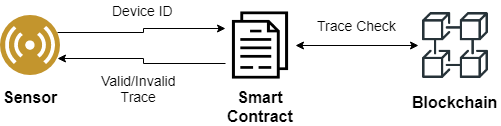}
  \caption{Sensor Traceability Verification using a Smart Contract}
  \label{fig:sensorcontract}
\end{figure}

Third, now that we have established what will be stored on the chain, we must understand how we can use the blockchain for traceability verification checks. Popular implementations, such as Ethereum, use smart contracts to execute code and interact directly with the blockchain. To perform traceability verification, within a secure boot process (i.e. when the sensor starts up), we can use a smart contract. The smart contract will execute code that will verify whether or not there is a complete traceability chain, with each unit in the chain having valid calibration, before the sensor is allowed to start capturing data (Figure~\ref{fig:sensorcontract}).

Specifically, the contract will take the sensor's device ID as input to the smart contract, which will execute a function to verify it has complete traceability. The algorithm, as shown in Algorithm~\ref{alg:getrootcert}, will retrieve the root (calibration) report of the device's traceability chain, which retrieves the parent report from the chain, verifies signatures, and loops until there are no parents. It will then check the final device's certifying organisation to see if it is an NMI, in our case NPL, and if so, the traceability chain is valid and complete, and thus return a {\em verified} result to the device. Likewise, if there is no NMI root, the trace will complete but will return a non-verified result to the device. Furthermore, to retrieve a certificate itself, the smart contract will interact with the PKI system (depicted in Figure~\ref{fig:calibrationhierarchy}) to retrieve the certificate. In the algorithm, the signatures will be verified before accepting the parent identifier. The public key of the technician who calibrated the device is not that of the one who signed the parent, then the verification will fail and return a null result (ultimately resulting in invalid traceability), and otherwise will continue looping until the NMI root.

\begin{algorithm}[H]
\begin{algorithmic}[1]
\footnotesize

  \Procedure{TraceCal{\_}READ}{$device\_id$}
	\State $device\_report = reports[device\_id]$
	\State $parent\_cert = certificates[device\_report.parent\_id]$
	\State $technician\_cert = certificates[device\_report.technician\_id]$

	\LineComment{If report is not signed by parent device, then fail}
	\If {$!(key\_verify(device\_report, parent\_cert))$}
		\State return $null$
	\EndIf

	\LineComment{If report is not signed by technician, then fail}
	\If {$!(key\_verify(device\_report, technician\_cert)$}
		\State return $null$
	\EndIf

	\State $org\_cert = certificates[technician\_cert.org\_id]$

	\If {$verify\_signature(technician\_cert, org\_cert) == false$}
		\State return $null$
	\EndIf

	\If {$check\_chain\_of\_trust(org\_cert, ROOT\_CERT) == false$}
		\State return $null$
	\EndIf

	\LineComment{Report now verified, now verify there is a root}

    \State $root\_report\_id = device\_id$
    \State $parent = reports[root\_report\_id].parent\_device$
    \LineComment{Loop until there is no parent}
    \While {$bytes(parent).length > 0$}
   		\LineComment{Verify parent report is signed by parent device}

    	\If {$key\_verify(parent,$\\ \hspace*{1.5cm}$certificates[parent].parent\_device)$}
    		\If {$key\_verify(parent, technician\_cert)$}
    			\State $root\_report\_id = parent$
        		\State $parent = reports[root\_report\_id].parent\_device$
    		\Else
    			\State return $null$
    		\EndIf
        \Else
        	\State return $null$
    	\EndIf
	\EndWhile
	\State return $reports[root\_report\_id]$
\EndProcedure
\normalsize
\end{algorithmic}
\caption{Trace Verification}
\label{alg:getrootcert}
\end{algorithm}

Fourth, upon calibration, the device will be imprinted with a ECDSA public and private keypair, which are signed by the certified technician, establishing a chain of trust. The technician's keys used to sign the device's calibration report and are in turn signed by the organisation's keys who certified the technician (Figure~\ref{fig:signing}), such that we can verify that the technicians themselves are not fake.

Certified calibration organisations will be associated with their own keypair, which will be used to sign all calibration technician keys they wish to certify.

%% file: sections/evaluation.tex
\section{Evaluation}
\label{sec:eval}

In order to better understand the natural consideration of blockchains as a solution to fulfil our protection requirements, we must evaluate an implementation that can verify the completeness of traceability chains at any stage in the calibration hierarchy.

\subsection{Blockchain Environment}

We used the Ethereum
blockchain~\cite{wood2014ethereum}, a Turing-complete, decentralised
{\em value-transfer} system which facilitates the use of smart
contracts, written in Solidity, to interact with the blockchain. The
programming language in Ethereum is implemented as a set of 140
opcodes which all nodes execute deterministically. The opcodes are
condensed to form a bytecode string which can be published on the
network, in the form of a smart contract. During deployment, a
transaction is created by the account deploying the contract, and the
contract is given its own unique address.  When this transaction is
accepted, the smart contract persists in the network. The contract may
have various functions, and can also allocate persistent memory on the
network. Any account which wants to interact with it uses the
contract's address to call its various functions. The contract may
contain two types of functions, including transactions and
calls. Transactions are those which modify the persistent memory of
the contract. They are called transactions specifically because they
need to be run by all nodes to ensure synchronicity, and thus cost
computational power. Calls merely read the persistent memory and can
be run locally as well, and hence are free of cost. Since each
transaction function requires computational resources based on the
bytecode executed, there must be a way to charge each operation. Thus,
every opcode is assigned a fixed cost which was tabulated when the
network was deployed, and this cost is measured by units of gas.

\subsubsection{Smart Contracts}

The purpose of our smart contract is to define the functions of our
protection mechanism, described in Section~\ref{sec:protmech}, to
interact with the blockchain to read and write data. In our smart
contract, we defined functions for declaring aspects of the
calibration hierarchy, as well as those for traceability
verification. Specifically, we defined functions for creating and
retrieving calibration organisations, certified calibration
technicians and calibration reports, as well as for conducting traceability
verification. Table~\ref{table:funcs} describes
the primary functions within our smart contract. Smart contracts were
deployed and tested on private Ethereum blockchain (Ganache) as well
as public Testnet (Ropsten Testnet) with the help of truffle testing
framework~\cite{truffle}.  An interesting fact is that the gas usage
is exactly same whether the smart contract is deployed on the Ethereum
Test Net (Ropsten) or a private Net (Ganache). After deployment of the
smart contracts we measured the gas usage of each function as well as
the transaction confirmation times.

\subsubsection{Calibration Hierarchy and Traceability}
To meet the definition of our protection mechanism, the smart contract should effectively be able to create the calibration hierarchy as well as provide methods to verify the metrological traceability of a given device or calibration unit. As previously described, we define functions to create calibration organisations and certified calibration technicians who will oversee and perform calibration of these devices, as well as providing a function for verifying traceability chains, {\em TraceCal{\_}READ}, described in detail in Algorithm~\ref{alg:getrootcert}.

\begin{table*}[h!]
  \begin{center}
    \begin{tabular}[t]{|c|*{2}{>{\raggedright\arraybackslash}p{0.65\linewidth}|}}
      \hline
      \textbf{Function} & \textbf{Description}\\
      \hline
      {\em createOrganisation} & Accepts an ID and a name, and creates an organisation on the blockchain \\
      {\em createTechnician} & Requests an Ethereum address and an organisation id, and will create a technician on the blockchain \\
      {\em createReport} & Accepts a number of parameters, such as the device id and technician id, and creates a calibration-report object on the blockchain \\
      {\em TraceCal{\_}READ} & Accepts a device ID and returns a root calibration-report if it has one, else returns the calibration-report of the device itself. \\
      {\em getParentReport} & Accepts a device and returns its direct parent's calibration-report or NULL.\\
      {\em getOrgName} & Accepts an organisation ID and returns the name of the organisation \\
      {\em getTechnicianOrganisation} & Returns the organisation ID of the organisation who certified the technician \\
      \hline
    \end{tabular}
  \end{center}

  \caption{List of Implementation Functions}
  \label{table:funcs}
\end{table*}

\vspace{-1.5cm}

\subsubsection{Ropsten Test Network}

In order to properly evaluate how our protection mechanism performs in a realistic environment, we deployed our implementation on the Ropsten test network~\cite{ropsten}. Also known as the Ethereum Testnet, the Ropsten test network is the largest Ethereum test network and runs the same Proof-of-Work (PoW) protocol as Ethereum, but is designed for testing smart contracts before deploying them on the main Ethereum network. It uses a form of Ether, Ethereum's currency, called {\em rEth} which costs no real money. However, this can also be produced from mining and can be received from faucets for testing transactions without imposing a legitimate cost.

In comparison with other Ethereum testnets such as Kovan~\cite{kovan} or Rinkeby~\cite{rinkeby}, which use an alternative Proof-of-Authority (PoA) consensus protocol and have lower block confirmation times, the PoW Ropsten testnet best reproduces the current Ethereum production environment conditions and is useful for testing our protection mechanism against realistic transaction rates/times, number of nodes/miners, and gas prices, compared to those on the main Ethereum network.

\subsection{Functionality Testing}

The aim of our first set of experiments was to determine whether or not our protection mechanism functions as intended. Specifically, our functions to set up organisations, technicians and calibration-reports must properly create their respective objects, with the appropriate input parameters, and raise errors when these parameters are invalid. 

\subsubsection{Creating organisations, technicians and reports}

When our smart contract is executed, we instantiate the calibration hierarchy with NPL as the root NMI organisation. From this, we tested creating organisations, with each certifying several technicians. These were created using the {\em createOrganisation} and {\em createTechnician} functions. These technicians would then go on to calibrate field devices and calibration units, which produces a calibration report upon completion of calibration; which ultimately need to be placed on the chain. To create a report, we use the {\em createReport} function in the smart contract. As expected, all our tests were successful, with the appropriate objects created on the chain. Appropriately, we also defined functions for data retrieval, such as {\em getTechnicianOrganisation} which gets the certifying organisation of a technician, for which all tests returned expected results. 

\subsection{Scalability Testing}

After developing the base of our system, we evaluated how our protection mechanism scales with the ubiquitous nature and vast size of the calibration hierarchy. As well as this, we also evaluated how the addition of signatures, used for signing calibration reports (among others as described in Section~\ref{sec:protmech}), affects how well our protection mechanism scales. The following tests which involve contract executing times have been run on a local blockchain using Ganache as the provider, and the Remix IDE to run the contract calls. We use Ganache to get the contract executing time as the contract is executed immediately, whereas on the main Ethereum network other contracts may be executed in the same block and measuring the execution time would be difficult. Likewise, we measured gas cost in the following experiments using the Ropsten network as Ganache provides an environment for testing contracts without costs, whereas Ropsten imposes gas and Ether costs like the main network but for free.

\subsubsection{Impact of \#Devices on Execution Time for Traces}

For our first set of experiments, we measured the impact the number of devices in the calibration hierarchy has on the execution time of the smart contract for traceability verification. Firstly, to match the calibration hierarchy, we used varying numbers of field devices $n$ as a baseline. From this, we deduce the number of levels as $log(n)$, such that if we have $100$ field devices, the calibration hierarchy will consist of two levels as well as the root NMI. Furthermore, we map the number of organisations in the calibration hierarchy as $log_2(n-1)$, such that for $100$ field devices there will be $4$ organisations.

\begin{figure}[ht!]
 \centering
  	\begin{subfigure}{.5\textwidth}
 		  \centering
 		  \includegraphics[width=.9\linewidth]{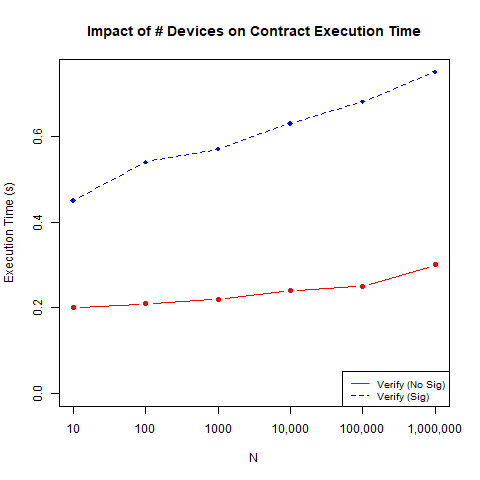}
   		\caption{Impact of \# Devices on Execution Time}
  		\label{fig:tracertimes}
	\end{subfigure}%
	\begin{subfigure}{.5\textwidth}
    \centering
  	\includegraphics[width=.9\linewidth]{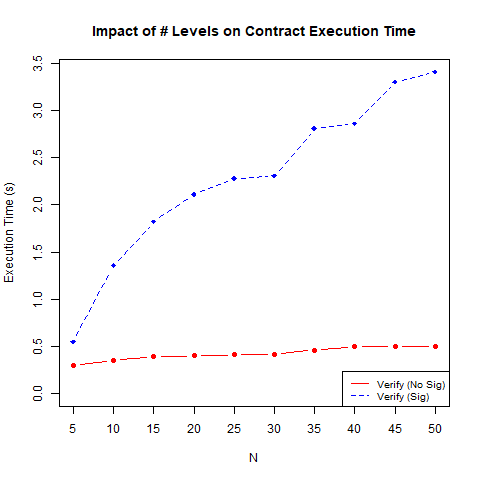}
  	\caption{Impact of \# Levels on Execution Time}
  	\label{fig:levelstime}
	\end{subfigure}
	\caption{Impact of \# Devices and \# Levels on Execution Time}
\end{figure}

Next, we define the scope of our first set of experiments for $n$ in
the range $10 \leq n \leq 10^6$. As shown in
Figure~\ref{fig:tracertimes}, we observed the effect of $n$ field
devices on the contract execution time for
verifying (read) traces. We observed that as the number of field
devices and levels increase, the time for verifying traces
increases.

Furthermore, we also observed the impact of adding signatures in our protection mechanism. For verification, the addition of signatures in the verification of traces more than doubles the contract execution time. Although this may seem a lot, if we consider the case of $1,000,000$ devices, the execution times are still only just over a single second. If we compare these times to what we would expect from the paper-based current state-of-the-art, they are an extremely significant improvement.

\subsubsection{Impact of \#Levels on Execution Time}

In our previous experiments, we derived the number of levels based on the number of field devices, $n$, as the primary variable. Realistically, there may be more than $log(n)$ levels, and it is interesting to evaluate how the number of levels impacts the execution time. In this experiment, we measured the impact of the number of levels on execution times for verifying traces. As shown in the results of this experiment in Figure~\ref{fig:levelstime}, the contract execution time increases in all cases as the number of levels increases. As the verification function requires reaching the root certificate, the time spent for will increase as the number of antecedent units in a device's trace also increases.
With respect to contract execution time, the addition of signatures seems to have little impact with a small number of levels, but increases significantly as the number of levels increase, taking around 7 seconds to create traces at 50 levels.

%% file: sections/discussion_ryan.tex
\section{Discussion}
\label{sec:discussion}

In safety-critical environments, it is vital that we ensure safe operation and robustness under adversarial pressure, during activities such as performing surgical procedures on humans. In Section~\ref{sec:requirements}, we have motivated a number of protection requirements, namely integrity, availability, anonymity and tamper-resistant storage.

\subsection{System Characteristics}

\paragraphb{Latency:} We demonstrate that traceability verification, at our worst case of 50 levels including signature verification in place, can be carried out in under 4 seconds. If we assume a typical surgical robot to realistically employ tens of sensors, such as the Raven II master device which makes use of 9 sensors for orientation and force-feedback alone~\cite{hannaford2012raven}, we can expect the verification process for all to complete within several minutes, roughly the time it takes for IV insertion, anaesthesia preparation, etc. during a typical pre-op procedure. 
As described by Hackel et al.~\cite{hackel2017digital}, NMIs at the root level will perform around 10,000 calibrations per year, intermediares ~100,000 per year and internal facilities situated in safety-critical environments to perform potentially millions of calibrations per year. Given the time to complete a traceability check, a key component of calibration itself, and the increasing numbers of IoT systems being implemented, the issues facing the current state-of-the-art will be evermore present.
The current state-of-the-art, being a collection of systems and processes dealing with manual paper records~\cite{hackel2017digital}, takes in the order of several hours to days for completing a single calibration verification check, in contrast to our system.

\paragraphb{Scalability:} Our evaluation involved testing our solution against a scaling calibration hierarchy. Specifically, our scaling factors included the number of field devices and the number of levels in the calibration hierarchy. In our case for on-the-fly calibration verification, which would happen at critical stages in system operation such as the secure boot, our system would need to successfully carry out the verification checks in quick succession as to minimise any delay in operation. For example, if we consider an emergency room scenario where a patient has been rushed into a trauma bay, these checks should be carried out and completed ready for reliable operation before the patient has reached the room. In our tests, we have demonstrated that even with 50 levels and the added use of signatures, the verification process takes under four seconds. Realistically, the trauma team in the scenario described would be notified before the patient enters the ER, and thus our solution therefore meets this criteria whilst maintaining resilience to malicious actors.


\paragraphb{Usability:} Naturally, one may question the necessity of storing the result of trace verifications (once confirmed) on the chain. Rather than repeating the full traceability verification, one could store the result (valid or invalid) of the trace on the chain itself, such that we can just check this result instead. As this is a write operation, using the Ethereum blockchain imposes a gas cost, so while it may be faster, it is important to assess the cost of writing trace results to the chain. The Ethereum network currently prescribes that each block may only contain transactions whose total gas cost amounts to $8,000,000$ gwei at maximum. This is not only a result of performance and financial considerations on the part of the nodes, but the need to minimize the computation required by non nodes with low computing power to validate blocks as well. The gas cost for this operation is approximately $200,000$ gwei per execution, and thus assuming a fairly strict regime of hourly verification checks would result in an approximate cost of $4,000,000$ gwei per device. Unfortunately, taking into account mining time and the Ethereum network, only 2 devices can be validated per block, assuming that the system has access to the entire global bandwidth. At the time of writing, the average mining time for Ethereum is \~ 15 seconds, and thus a set of fully harmonised trace write operations is limited to around $11,520$ per day. Therefore, using the Ethereum blockchain for this write operation in particular does not seem viable, in terms of applicability as well as economically. However, other blockchain technologies may be better suited to enabling this, a point for future work.

\subsection{Security and Privacy}
We show that a decentralised blockchain has significant enabling power as a collaboration mechanism between various players involved in ensuring calibration of safety critical systems  -- OEMs, users, and third-party calibration agencies, and NMIs. In particular, the use of a permissionless blockchain helps address privacy concerns over user profiling by internal adversaries who can apply traffic analysis techniques over calibration traffic (the anonymity property). Combining blockchains with appropriate certificate management prevents malicious action by internal and external adversaries to resist fake calibration reports, fake technicians. The ability to ensure that all sensor measurements can be verified for correctness prior to use together lays the foundation for secure safety-critical systems. Additionally, it enables transparency in the supply chain which may have  positive secondary effects, such as extending the collaboration within the supply-chain to support device provenance among other applications.

We now discuss how our solution meets various security properties.

\paragraphb{System Integrity:} Tampering with calibration records, is prevented by the inherent properties of blockchains, specifically the chaining of data and the consensus mechanism in place. Calibration integrity -- proving that a device has valid calibration -- for example that it has not been revoked ({\em fresh}), is met through traceability verification. Furthermore, the forgery of calibration reports is prevented by having the certified calibration technician sign the reports for the devices they calibrate.

\paragraphb{System Forensics:} The system as a whole needs to be able to withstand hostile scrutiny in a court of law. Several cases~\cite{lawsuitsurgical,davincilawsuit} are clear demonstrations of the potential liabilities involved and the stakes revolving around safety within IoT contexts. Support for system forensics are required to ensure that one can successfully verify that calibration was in fact carried out, and to support this a record of such events leading to verification must be recorded to be verified at a later date. In comparison with the current state-of-the-art, where forensics is carried out over disconnected centralised databases of calibration reports, our blockchain-based system keeps a strong, tamper-resistant trail of evidence that can be followed throughout entire traceability chains. Furthermore, we note that, albeit not part of our evaluation, additional smart contracts can be written to support generalised and feature-specific forensic and auditing applications.

\paragraphb{Anonymity:} To prevent information leakage to other parties in the chain, all read operations are anonymous as the data is read from the verifier's local machine, and thus prevents the learning of links between the verifier themselves the device being traced. Using the Ethereum platform for our solution, being a public permissionless blockchain, readers and verifiers of the stored data are kept anonymous, unlike writing to the chain which is not. As described in Section~\ref{sec:eval}, our solution makes use of only read operations in terms of traceability verification, where anonymity is required.

\paragraphb{Availability:} Currently, with the use of centralised data storage for calibration records leaves the calibration ecosystem vulnerable to targeted attacks on and compromise of the storage servers. Thus, the consideration over the secure design of the storage infrastructure is important, with delays or denials of verification potentially heavily impacting operational efficiency and accuracy. Simply, a decentralised storage infrastructure can meet the requirement for high availability and scale, additional mechanisms would need to be coordinated to meet the rest of our requirements, to which the blockchain solution fortunately provides.

%% file: sections/conclusion.tex
\section{Conclusion}
\label{sec:conclusion}

An open challenge within industrial IoT processes is maintaining the
integrity of calibration under adversarial pressure. Whilst there are
many factors which contribute to this, including software patches to
secure data storage, an important foundational requirement is to secure
the calibration mechanism itself. In particular, the need for a mechanism
that is: highly available, verifiable and tamper-resistant, for verifying
traceability is becoming clear. In our research, we propose
a mechanism that successfully establishes traceability chains, to ensure
we can maintain valid calibration and rapidly attend to adversarial faults,
leveraging blockchains as a highly-available tamper-resistant chain of evidence.

%% file: paper.bbl
\begin{thebibliography}{10}
\providecommand{\url}[1]{\texttt{#1}}
\providecommand{\urlprefix}{URL }
\providecommand{\doi}[1]{https://doi.org/#1}

\bibitem{alicke2017supply}
Alicke, K., Rexhausen, D., Seyfert, A.: Supply chain 4.0 in consumer goods
  (2017)

\bibitem{de2000calibration}
de~Castro, C.N., Louren{\c{c}}o, M., Sampaio, M.: Calibration of a dsc: its
  importance for the traceability and uncertainty of thermal measurements.
  Thermochimica Acta  \textbf{347}(1-2),  85--91 (2000)

\bibitem{ropsten}
Foundation, E.: Ropsten testnet pow chain.
  \url{https://github.com/ethereum/ropsten}

\bibitem{truffle}
Group, T.B.: Sweet tools for smart contracts | truffle suite.
  \url{https://www.trufflesuite.com/}

\bibitem{hackel2017digital}
Hackel, S., H{\"a}rtig, F., Hornig, J., Wiedenh{\"o}fer, T.: The digital
  calibration certificate. Metrologie f{\"u}r die Digitalisierung von
  Wirtschaft und Gesellschaft pp. 75--82 (2017)

\bibitem{hannaford2012raven}
Hannaford, B., Rosen, J., Friedman, D.W., King, H., Roan, P., Cheng, L.,
  Glozman, D., Ma, J., Kosari, S.N., White, L.: Raven-ii: an open platform for
  surgical robotics research. IEEE Transactions on Biomedical Engineering
  \textbf{60}(4),  954--959 (2012)

\bibitem{huang2005computer}
Huang, S.H., Sheoran, S.K., Keskar, H.: Computer-assisted supply chain
  configuration based on supply chain operations reference (scor) model.
  Computers \& Industrial Engineering  \textbf{48}(2),  377--394 (2005)

\bibitem{jcgm2012200}
JCGM, J.: 200: 2012—international vocabulary of metrology—basic and general
  concepts and associated terms (vim). Tech. rep., Technical Report (2012)

\bibitem{kaarls1997comite}
Kaarls, R., Quinn, T.: The comit{\'e} consultatif pour la quantit{\'e} de
  mati{\`e}re: a brief review of its origin and present activities. metrologia
  \textbf{34}(1), ~1 (1997)

\bibitem{kovan}
Kovan: Kovan - stable ethereum public testnet.
  \url{https://github.com/kovan-testnet/proposal}

\bibitem{leach1999calibration}
Leach, R.K.: Calibration, traceability and uncertainty issues in surface
  texture metrology.  (1999)

\bibitem{lu2017industry}
Lu, Y.: Industry 4.0: A survey on technologies, applications and open research
  issues. Journal of Industrial Information Integration  \textbf{6},  1--10
  (2017)

\bibitem{lukavc2015fourth}
Luka{\v{c}}, D.: The fourth ict-based industrial revolution" industry
  4.0"—hmi and the case of cae/cad innovation with eplan p8. In:
  Telecommunications Forum Telfor (TELFOR), 2015 23rd. pp. 835--838. IEEE
  (2015)

\bibitem{pereira2016uncertainty}
Pereira, P.: Uncertainty of measurement in medical laboratories. In: New Trends
  and Developments in Metrology. InTech (2016)

\bibitem{quarta2017experimental}
Quarta, D., Pogliani, M., Polino, M., Maggi, F., Zanchettin, A.M., Zanero, S.:
  An experimental security analysis of an industrial robot controller. In: 2017
  IEEE Symposium on Security and Privacy (SP). pp. 268--286. IEEE (2017)

\bibitem{rinkeby}
Rinkeby: Rinkeby: Ethereum testnet. \url{https://www.rinkeby.io/}

\bibitem{roblek2016complex}
Roblek, V., Me{\v{s}}ko, M., Krape{\v{z}}, A.: A complex view of industry 4.0.
  Sage Open  \textbf{6}(2),  2158244016653987 (2016)

\bibitem{lawsuitsurgical}
Siegel, E.R., McFadden, C., Monahan, K., Lehren, A.W., Siniauer, P.: The da
  vinci surgical robot: A medical breakthrough with risks for patients.
  \url{https://www.nbcnews.com/health/health-news/da-vinci-surgical-robot-medical-breakthrough-risks-patients-n949341}
  (12 2018)

\bibitem{talamini2003prospective}
Talamini, M.A., Chapman, S., Horgan, S., Melvin, W.S.: A prospective analysis
  of 211 robotic-assisted surgical procedures. Surgical Endoscopy and Other
  Interventional Techniques  \textbf{17}(10),  1521--1524 (2003)

\bibitem{davincilawsuit}
Watch, S.: da vinci robot lawsuit.
  \url{http://surgicalwatch.com/davinci-robot/lawsuit/} (2015)

\bibitem{wood2014ethereum}
Wood, G., et~al.: Ethereum: A secure decentralised generalised transaction
  ledger. Ethereum project yellow paper  \textbf{151}(2014),  1--32 (2014)

\end{thebibliography}
